\documentstyle[12pt,a4,psfig]{article}

\def\vsk#1{\noalign{\vskip#1 cm}}
\def\vsp#1{\vspace{#1 cm}}
\def\ov{\overline}
\def\disp{\displaystyle}
\def\gsim{{\mathop >\limits_\sim}}
\def\lsim{{\mathop <\limits_\sim}}

\def\su321 {${\rm SU(3)}_C\times {\rm SU(2)}_L\times {\rm U(1)}_Y$\ }

\def\ky{k_Y^{}}

\def\gev{{\rm GeV} }

\def\gm5{\gamma_5}

\def\mmz{m_Z^2}

\def\pibar{\overline{\Pi}}

\def\xs{x_s}
\def\xa{x_\alpha}
\def\munif{m_U^{}}
\def\au{\alpha_U}
\def\mt{m_t^{}}
\def\xt{x_t^{}}
\def\xh{x_H^{}}
\def\mh{m_H^{}}
\def\mz{m_Z^{}}
\def\msbar{$\ov{{\rm MS}}$}

\def\ebar{\bar{e}^2}
\def\sbar{\bar{s}^2}

\def\abar{\bar{\alpha}}

\def\yhat{\hat{y}}
\def\ahat{\hat{\alpha}}
\def\ehat{\hat{e}}

\def\ghat{\hat{g}}

\def\etal{{\it et al.}}
\newcommand{\beq}{\begin{equation}}
\newcommand{\eeq}{\end{equation}}
\newcommand{\bea}{\begin{eqnarray}}
\newcommand{\eea}{\end{eqnarray}}
\newcommand{\bsub}{\begin{subequations}}
\newcommand{\esub}{\end{subequations}}

\def\PRD#1#2#3{Phys. Rev. {\bf D#1} (19#2) #3}
\def\NPB#1#2#3{Nucl. Phys. {\bf B#1} (19#2) #3}
\def\ZPC#1#2#3{Z. Phys. {\bf C#1} (19#2) #3}
\def\PLB#1#2#3{Phys. Lett. {\bf B#1} (19#2) #3}
\def\PRL#1#2#3{Phys. Rev. Lett. {\bf #1} (19#2) #3}
\makeatletter
\newtoks\@stequation

\def\subequations{\refstepcounter{equation}%
  \edef\@savedequation{\the\c@equation}%
  \@stequation=\expandafter{\theequation}
  \edef\@savedtheequation{\the\@stequation}
  \edef\oldtheequation{\theequation}%
  \setcounter{equation}{0}%
  \def\theequation{\oldtheequation\alph{equation}}}

\def\endsubequations{%
  \ifnum\c@equation < 2 \@warning{Only \the\c@equation\space subequation
    used in equation \@savedequation}\fi
  \setcounter{equation}{\@savedequation}%
  \@stequation=\expandafter{\@savedtheequation}%
  \edef\theequation{\the\@stequation}%
  \global\@ignoretrue}

\def\eqnarray{\stepcounter{equation}\let\@currentlabel\theequation
\global\@eqnswtrue\m@th
\global\@eqcnt\z@\tabskip\@centering\let\\\@eqncr
$$\halign to\displaywidth\bgroup\@eqnsel\hskip\@centering
     $\displaystyle\tabskip\z@{##}$&\global\@eqcnt\@ne
      \hfil$\;{##}\;$\hfil
     &\global\@eqcnt\tw@ $\displaystyle\tabskip\z@{##}$\hfil
   \tabskip\@centering&\llap{##}\tabskip\z@\cr}

\makeatother

\begin{document}
\thispagestyle{empty}
\vspace*{-15mm}
\baselineskip 10pt
\begin{flushright}
\begin{tabular}{l}
{\bf KEK-TH-537}\\
{\bf hep-ph/9709279}\\
November 1997
\end{tabular}
\end{flushright}
\baselineskip 18pt 
\vglue 15mm 

\begin{center}
{\Large\bf
String unification scale and the hyper-charge Kac-Moody 
level in the non-supersymmetric standard model
}
\vspace{5mm}

\def\thefootnote{\alph{footnote}}
\setcounter{footnote}{0}

{\bf
Gi-Chol Cho$^{1,}$\footnote{Research Fellow of the Japan Society 
for the Promotion of Science} and 
Kaoru Hagiwara$^{1,2}$
}

\vspace{5mm}
\def\thefootnote{\arabic{footnote}}
\setcounter{footnote}{0}

$^1${\it Theory Group, KEK, Tsukuba, Ibaraki 305, Japan}\\
$^2${\it ICEPP, University of Tokyo, Hongo, Bunkyo-ku, 
Tokyo 113, Japan}

\vspace{20mm}
\end{center}


\begin{center}
{\bf Abstract}\\[10mm]
\begin{minipage}{12cm}
\noindent
The string theory predicts the unification of the gauge 
couplings and gravity.  The minimal supersymmetric 
Standard Model, however, gives the unification scale 
$\sim 2\times 10^{16}$~GeV which is significantly 
smaller than the string scale $\sim 5\times 10^{17}$~GeV 
of the weak coupling heterotic string theory.  
We study the unification scale of the non-supersymmetric 
minimal Standard Model quantitatively at the two-loop level.  
We find that the unification scale should be at most 
$\sim 4\times 10^{16}$~GeV and the desired Kac-Moody level 
of the hyper-charge coupling should be $1.33~ \lsim~ \ky~ 
\lsim~ 1.35$. 
\end{minipage}
\end{center}
%
\newpage
%
\baselineskip 20pt
The theory of $E_8 \times E_8$ heterotic string~\cite{hetero} 
has some attractive impacts on the model of low-energy 
particle physics.  
The theory has a potential of explaining the low-energy 
gauge groups, the quantum numbers of 
quarks, leptons and the Higgs bosons, the number of generations, 
and the interactions among these light particles which are 
not dictated by the gauge principle.  
One of the immediate consequences of the string theory is 
the unification of the gauge interactions and the gravity.  
Since, in the string theory, gravitational and gauge 
interactions are naturally related, the strength of the 
gauge couplings and the unification scale are both given by 
the Newton constant.  
The unification scale of the heterotic string theory is 
predicted to be~\cite{string_unif, dienes} 
\beq 
\munif|_{\rm string} \approx 5 \times 10^{17}~{\rm GeV}, 
\label{eq:mx_string}
\eeq
in the weak coupling limit where the 1-loop string 
effects are taken into account.  On the other hand, the minimal 
supersymmetric Standard Model (MSSM) predicts the unification 
scale 
\beq
\munif|_{\rm MSSM} \approx 2 \times 10^{16}~{\rm GeV}, 
\label{eq:mx_mssm}
\eeq
by using the recent results of precision electroweak 
measurements as inputs.
The discrepancy between (\ref{eq:mx_string}) and 
(\ref{eq:mx_mssm}) is a few percent of the logarithms of 
these scales. 
However the extrapolation of (\ref{eq:mx_string}) to 
the weak scale leads the experimentally unacceptable values 
of $\sin^2\theta_W$ and $\alpha_s$ under the hypothesis that 
the spectrum below the string scale is that of the MSSM. 
Various attempts to modify this naive prediction are reviewed 
in ref.~\cite{dienes}. 
For instance, the 2-loop string effects are not known. 
On the other hand, it has been suggested~\cite{witten} that the 
strong coupling limit of the $E_8 \times E_8$ heterotic 
string theory, which is considered to be the 
11-dimensional M-theory, can give rise to a significantly 
lower string scale than the estimation (\ref{eq:mx_string}) 
in the weak coupling limit. 

Alternatively, the gauge coupling unification scale can be 
modified in string theories with non-standard Kac-Moody 
levels. 
The coupling constant $g_U$, which is related to the Newton 
constant in the string theory, is expressed in terms of 
the SU(3)$_C$, SU(2)$_L$ and U(1)$_Y$ gauge couplings 
and the corresponding Kac-Moody level $k_i~(i=Y,2,3)$ 
as~\cite{ginsparg}
\beq
g_U^2 = k_i g_i^2, 
\eeq
at the unification scale $\munif$. 
The factor $k_i$ should be positive integer for the non-Abelian 
gauge group. 
On the other hand, for the Abelian group, its value depends on 
the structure of four-dimensional string models. 
In view of the gauge field theory, $k_i$ plays the role of a 
normalization factor for $g_i$ and, for example, 
the set $(\ky, k_2, k_3) = (5/3,1,1)$ is taken to embed 
the hyper-charge $Y$ in the SU(5) GUT group.

It has been known that the SU(5) grand unification is not achieved 
if one extrapolates the observed three gauge couplings by using the 
renormalization group equations (RGE) in the minimal Standard 
Model (SM). 
It has been noted~\cite{dienes}, however, that 
the trajectories of the SU(2)$_L$ and the SU(3)$_C$ couplings 
intersect at near the unification scale $\munif$ predicted by the 
string theory: 
for example, the leading order RGE with a certain 
choice of the weak mixing angle and the QED coupling in the 
\msbar\ scheme, 
\bsub
\bea
\sin^2\theta_W(\mz)_{\ov{{\rm MS}}} &=& 0.2315, \\
1/\alpha(\mz)_{\ov{{\rm MS}}} &=& 128 , 
\eea
\esub
gives the following results, 
\bsub
\bea
\munif &\approx& 1\times 10^{17}~{\rm GeV}~~~~~~~~~~~~~
{\rm for}~~ \alpha_s(\mz) = 0.118,\\
    &\approx& 2 \times 10^{17}~{\rm GeV}~~~~~~~~~~~~~
{\rm for}~~ \alpha_s(\mz) = 0.121.
\eea
\label{mu_sm_1loop}
\esub
\vsp{-0.4}
\\
The above unification scale $\munif$ is remarkably close to 
the string scale (\ref{eq:mx_string}), which may suggest 
the string unification without supersymmetry for the 
Kac-Moody level $\ky \approx 1.27$ for $k_2 = k_3 = 1$. 

Of course, deserting supersymmetry (SUSY) after compactification 
into four-dimension means that both the gauge hierarchy and the 
fine-tuning problems have to be solved without SUSY. 
The existence of a consistent string theory without the 
four-dimensional SUSY has not been demonstrated. 
It has been argued that the solution to these problems, 
if it exists, should be intimately related to the vanishing 
of the cosmological constant; see, $e.g.$, ref.~\cite{dienes} 
for a review of some exploratory investigations. 
Recently, as an application of this idea of minimal particle 
contents, the mechanism of baryogenesis in non-SUSY, 
non-GUT string model has been proposed~\cite{aoki}. 

In this letter we examine quantitatively at the 
next-to-leading-order (NLO) level the possibility of the 
string unification of the gauge couplings in the SM 
without SUSY. 
Because, in the string theory, the U(1)$_Y$ coupling can be 
rather arbitrarily normalized by the Kac-Moody level $\ky$,  
we define $\munif$ as the scale at which the trajectories 
of the SU(2)$_L$ and the SU(3)$_C$ running couplings intersect 
with $k_2 = k_3 = 1$\footnote{No attractive solution is found for 
$k_2 \neq k_3$.}. 
Our purposes are to find the scale $\munif$ and the corresponding 
$\ky$ under the current experimental and theoretical constraints 
on the parameters in the minimal SM. 
In the NLO level, the scale $\munif$ is not only affected by 
the uncertainty in the SU(3)$_C$ coupling but also by threshold 
corrections due to the SM particles such as the top-quark and 
the Higgs boson.  
The top-quark Yukawa coupling affects the RGE at the two-loop 
level.
Therefore, it is interesting to examine whether the scale 
$\munif$ in the minimal SM 
still lie in the string scale $\sim O(10^{17}~\gev)$ 
after the NLO effects are taken into account.

We first evaluate quantitatively the U(1)$_Y$ and SU(2)$_L$ 
\msbar\ couplings at the weak scale boundary of the RGE. 
The magnitudes of the \msbar\ couplings are determined in 
general by comparing the perturbative expansions of a certain 
set of physical observables with the corresponding experimental 
data. 
The correspondence can be made manifest by using 
the effective charges $\ebar(q^2)$ and $\sbar(q^2)$ of 
ref.~\cite{hhkm}.
The \msbar\ couplings $\ahat(\mu)= \ehat^2(\mu)/4\pi $ 
and $\ahat_2(\mu)= \ghat_2^2(\mu)/4\pi$ 
are related with the effective 
charges as  
\bsub 
\bea
\frac{1}{\abar(q^2)} &=& \frac{1}{\ahat(\mu)}  
+ 4\pi {\rm Re} \pibar^{QQ}_{T,\gamma}(\mu; q^2), 
\label{bar_ms_e2}\\
\frac{\sbar(q^2)}{\abar(q^2)} &=& \frac{1}{\ahat_2(\mu)} 
+ 4\pi{\rm Re} \pibar^{3Q}_{T,\gamma}(\mu; q^2),
\label{bar_ms_g2}
\eea
\label{bar_ms}
\esub \vsp{-0.6}
\\
where $\abar(q^2) = \ebar(q^2)/4\pi$. 
The explicit form of the vacuum polarization functions 
$\pibar^{AB}_{T,V}(\mu; q^2)$ in the SM 
can be found in Appendix A of ref.~\cite{hhkm}.
The above expressions are manifestly RG invariant in the 
one-loop order and give good perturbative expansions at 
$q^2 = m_Z^2$ for $\mu = \mz$. 
We hence need as inputs $\abar(m_Z^2)$ and $\sbar(m_Z^2)$.
Recent estimate of the hadronic contribution to the 
running of the effective QED charge finds~\cite{eidelman} 
\beq
1/\abar(m_Z^2) = 128.75 \pm 0.09. 
\label{QED_charge}
\eeq
All the other recent estimations~\cite{delta_h} find 
consistent results. 
Relation between the running QED charge of 
refs.~\cite{eidelman, delta_h} 
and the effective charge $\abar(q^2)$ of ref.~\cite{hhkm} 
that contain the $W$-boson contribution is found in 
ref.~\cite{hhm}. 
The effective charge $\sbar(m_Z^2)$ is measured directly 
at LEP1 and SLC from various asymmetries on the 
$Z$-pole~\cite{hhkm, hhm}. 
In the SM, however, its magnitude can be accurately calculated 
as a function of $\mt$ and $\mh$ through 
the following formula~\cite{hhkm, hhm}, 
\begin{eqnarray}
\sbar(m_Z^2) 
&=& 
\frac{1}{2} - 
\sqrt{
\frac{1}{4} - 4\pi\ov{\alpha}(m_Z^2) 
\biggl( 
\frac{1+0.0055-\alpha T}
{4\sqrt{2}G_F m_Z^2}
+ \frac{S}{16\pi}
\biggr)
      }, 
\end{eqnarray}
where $G_F$ and $\alpha$ are the Fermi coupling constant and the 
fine structure constant, respectively.
Accurate parametrizations of the SM contributions to the 
$S$ and $T$ parameters~\cite{stu} are found in ref.~\cite{hhm}, 
as functions of the scaled mass parameters 
\bsub
\bea
\xt &\equiv& \frac{\mt(\gev)-175~\gev}{10~\gev}, \\
\vsk{0.2} 
\label{top_param}
\xh &\equiv& \log \frac{\mh(\gev)}{100\ {\rm GeV}}. 
\eea
\esub 
Finally the \msbar\ coupling of the effective 5-quark QCD has 
been estimated as~\cite{pdg96} 
\beq
\alpha_s(\mz) = 0.118 \pm 0.003. 
\label{alphas_5q}
\eeq 
For later convenience, we introduce the following parametrizations 
to the observed and calculated values of the three effective 
charges of the SM: 
\bsub
\bea
\frac{1}{\abar(m_Z^2)} &=& 128.75 + 0.09 \xa, \\
\vsk{0.1}
\frac{\sbar(m_Z^2)}{\abar(m_Z^2)} &=& 29.66 - 0.044 \xt 
		+ 0.067 \xh + 0.002 x_H^2 - 0.01 \xa, 
\label{abar_2}\\
\alpha_s(\mz) &=& 0.118 + 0.003 \xs, 
\eea
\esub
where $\xs$ and $\xa$ are defined as 
\bsub
\bea
\xs &\equiv& (\alpha_s(\mz) - 0.118)/0.003, \\
\xa &\equiv& (1/\abar(m_Z^2) - 128.75)/0.09. 
\eea
\esub
The three \msbar\ couplings of the SM that enter as the 
boundary condition of the 2-loop RGE are then determined 
via eqs.~(\ref{bar_ms}) and the corresponding matching equation 
of the 5-quark and 6-quark QCD as follows: 
\bsub
\bea
\frac{\ky}{\ahat_1(\mz)} &=& \frac{1-\sbar(m_Z^2)}{\abar(m_Z^2)} 
- 0.77 + 0.19 \log \frac{\mt}{\mz}, 
\label{a1_ms}\\
\frac{1}{\ahat_2(\mz)} &=& \frac{\sbar(\mmz)}{\abar(m_Z^2)}
- 0.11 + 0.12 \log\frac{\mt}{\mz},
\label{a2_ms}\\
\frac{1}{\ahat_3(\mz)} &=& \frac{1}{\alpha_s(\mz)} + \frac{1}{3\pi}
\log\frac{\mt}{\mz}.
\label{a3_matching}
\eea
\esub
We use (\ref{a1_ms}) to (\ref{a3_matching}) as inputs to determine 
the unification scale $\munif$, and the relation $\ahat_1(\mu) = 
\ky \ahat_Y(\mu)$ to fix the desired Kac-Moody level $\ky$. 

The estimates (\ref{QED_charge}) and (\ref{alphas_5q}) give, 
respectively, $\xa = 0 \pm 1$ and $\xs = 0 \pm 1$. 
The observed top-quark mass~\cite{mt96} 
$\mt = 175 \pm 6$ GeV gives $\xt = 0 \pm 0.6$. 
The global fit including the electroweak precision experiments 
gives~\cite{hhm} 
$\mt = 172 \pm 6$ GeV, or $\xt = -0.3 \pm 0.6$. 
The error estimate of eq.~(\ref{QED_charge}) is 
conservative~\cite{hhm}, while that of eq.~(\ref{alphas_5q}) 
may be too optimistic. 
We will therefore explore the region of $|\xa|~\lsim~1,~ 
|\xt|~\lsim~1$ and $|\xs|~\lsim~2$. 
As for the Higgs boson mass $\mh$, the measurements of 
$\sbar(m_Z^2)$ and the other electroweak observables constrain 
it indirectly~\cite{hhm}, while the direct search at LEP gives 
$\mh~\gsim~70~\gev$. 
In addition, there are theoretical bounds, both the lower 
and the upper limits in order for the minimal SM to be valid up 
to the unification scale $\munif$.
The lower limit of $\mh$ is obtained from the stability of 
the SM vacuum. 
Its recent evaluation~\cite{altarelli_isidori, casas} finds 
\beq
\mh > 137.1 + 21 \xt + 2.3 \xs~~\gev
~~~~~~{\rm for~\Lambda \sim 10^{19}~GeV}. 
\label{higgs_low}
\eeq
Since the dependence on the cut-off scale $\Lambda$ is found 
to be small for $\Lambda > 10^{15}~\gev$~\cite{altarelli_isidori}, 
we can adopt eq.~(\ref{higgs_low}) as the lower limit 
of $\mh$ for $\Lambda \sim \munif$.
On the other hand, 
the upper bound is obtained by requiring the effective Higgs 
self-coupling to remain finite up to the cut-off scale $\Lambda$. 
A recent study finds~\cite{lee_kim};
\beq
\mh < 260 \pm 10 \pm 2~~\gev
~~~~~~{\rm for~\Lambda \sim 10^{15}~GeV}, 
\label{higgs_upp}
\eeq
where the first error denotes the uncertainty of theoretical 
estimation and the second one comes from the experimental uncertainty 
in $\mt$. 
Since the $\mt$-dependence of the upper limit is rather small, and 
since the upper limit decreases as $\Lambda$ increases, 
we set the upper limit of $\mh$ to be 
$270~\gev$ for $\Lambda \sim \munif$.
%
%
\begin{figure}[t]
\begin{center}
\leavevmode\psfig{figure=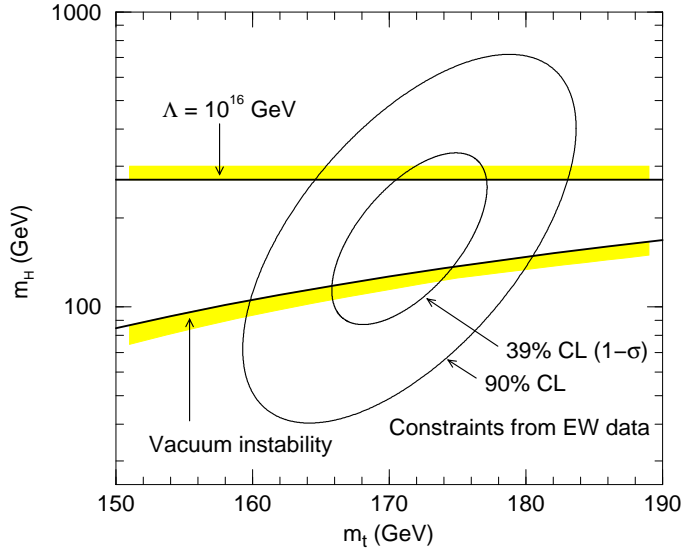,width=9cm}
\end{center}
\caption{ Constraint on the Higgs boson mass for the 
electroweak precision measurement and the theoretical 
bounds of the Higgs potential. 
The contours are obtained from the SM fit to all electroweak 
data with $\mt = 175 \pm 6$ GeV, $\alpha_s = 0.118 \pm 0.003$ 
and $1/\abar(m_Z^2) = 128.75 \pm 0.09$. 
The inner and outer contours correspond to 
$\Delta \chi^2 = 1$ $(\sim 39\%{\rm CL})$, and 
$\Delta \chi^2 = 4.61$ $(\sim 90\%{\rm CL})$, 
respectively~\cite{hhm}.
The upper and lower lines come from the triviality and vacuum 
stability bounds for the cut-off scale $\Lambda \sim 10^{16}$ GeV. 
}
\label{const_mh}
\vsp{0.5}
\end{figure}
In summary, we consider the following range of the Higgs boson 
mass 
\beq
137.1 + 21 \xt + 2.3 \xs < \mh~(\gev) < 270,
\label{mh_th}
\eeq
in our analysis.
We show in Fig.~\ref{const_mh} the allowed region of the Higgs 
boson mass which is obtained from the SM fit to all electroweak 
precision measurements~\cite{hhm}, where the contours are 
obtained from the SM fit to all the electroweak data and 
the external constraints $\mt = 175 \pm 6$ GeV~\cite{mt96}, 
$\alpha_s = 0.118 \pm 0.003$~\cite{pdg96} 
and $1/\abar(m_Z^2) = 128.75 \pm 0.09$~\cite{eidelman}.
Theoretically allowed region of $\mh$, eq.~(\ref{mh_th}), is 
also shown in the figure. 
It is clearly seen that the theoretically allowed range of 
$\mh$ with $\Lambda > 10^{16}~\gev$ is in perfect 
agreement with the constraint from these precision 
electroweak experiments. 

The 2-loop RGE for the gauge couplings 
$\ahat_i(\mu)$ in the \msbar\ scheme 
is given as follows; 
\beq
\mu \frac{d \ahat_i}{d \mu} = \frac{1}{2\pi} b_i 
\ahat_i^2 + \frac{1}{8\pi^2} \ahat_i^2 \biggl[ 
b_{ij}\ahat_j + c_{ik} \frac{\yhat_k^2}{4\pi} \biggr], 
\label{eq:rge_1}
\eeq
where $i=1,2,3$ and $k=t,b,\tau$. 
The U(1) hyper-charge normalization is taken as 
$\ahat_1 = \ky\ahat_Y$.
The term $\yhat_k$ denotes the \msbar\ Yukawa coupling.
The coefficients $b_i, b_{ij}$ and $c_{ik}$ are given 
in the minimal SM as~\cite{beta_sm}; 
\bsub
\begin{eqnarray}
b_i &=& \biggl(  \disp{\frac{1}{\ky}\frac{41}{6}, -\frac{19}{6}, 
-7 \biggr )}, \\
\vsk{0.4}
b_{ij} &=& \left ( 
  \begin{array}{ccc}
   \disp{\frac{1}{k_Y^2}\frac{199}{18}} 
  &\disp{\frac{1}{\ky}\frac{9}{2}}
  &\disp{\frac{1}{\ky}\frac{44}{3}} \\
\vsk{0.3}
   \disp{\frac{1}{\ky}\frac{3}{2}} & \disp{\frac{35}{6}} & 12 \\
\vsk{0.3}
   \disp{\frac{1}{\ky}\frac{11}{6}}& \disp{\frac{9}{2}} & -26 
   \end{array}
            \right ), \\
\vsk{0.4}
c_{ik} &=& \left ( 
  \begin{array}{ccc}
  \disp{-\frac{1}{\ky}\frac{17}{6}} & \disp{-\frac{1}{\ky}\frac{5}{6}} 
        & \disp{-\frac{1}{\ky}\frac{5}{2}} \\
\vsk{0.3}
  \disp{-\frac{3}{2}} & \disp{-\frac{3}{2}} & \disp{-\frac{1}{2}} \\
\vsk{0.3}
   -2 & -2 & 0
   \end{array}
            \right ).
\end{eqnarray}
\label{eq:rge_2}
\esub 
\vspace{-0.2cm}
\\
The \msbar\ Yukawa coupling for fermion $f$ 
is given in terms of the corresponding pole mass $m_f$ as 
\beq
\yhat_f(\mu) = 2^{3/4} G_F^{1/2} m_f \{ 1 + \delta_f(\mu)\}, 
\eeq
where the factor $\delta_f(\mu)$ denotes the QCD and electroweak 
corrections. 
Because only the top-quark Yukawa coupling is found to affect our 
results significantly, we set $\yhat_b = \yhat_\tau = 0$.
The explicit form of $\delta_t(\mu)$ has been given in ref.~\cite{kniehl}.
Only the leading order $\mu$-dependence of $\yhat_t(\mu)$ is needed 
in our analysis~\cite{beta_sm}; 
\bea
\mu \frac{d}{d\mu} \biggl( \frac{\yhat_t^2}{4\pi} \biggr) 
&=& 
\frac{1}{2\pi} \biggl( \frac{\yhat_t^2}{4\pi} \biggr) 
\left[ 
-\frac{1}{\ky}\frac{17}{12} \ahat_1 -\frac{9}{4} \ahat_2 
-8\ahat_3 + \frac{9}{2} \biggl( \frac{\yhat_t^2}{4\pi} \biggr) 
\right].
\eea
We can now solve the RGE in the NLO level and find 
the unification scale $\munif$ and the unified coupling $\au$ as 
functions of $\alpha_s(\mz)$, $\mt, \mh$ and $\abar(m_Z^2)$.

\begin{figure}[t]
\begin{center}
\leavevmode\psfig{figure=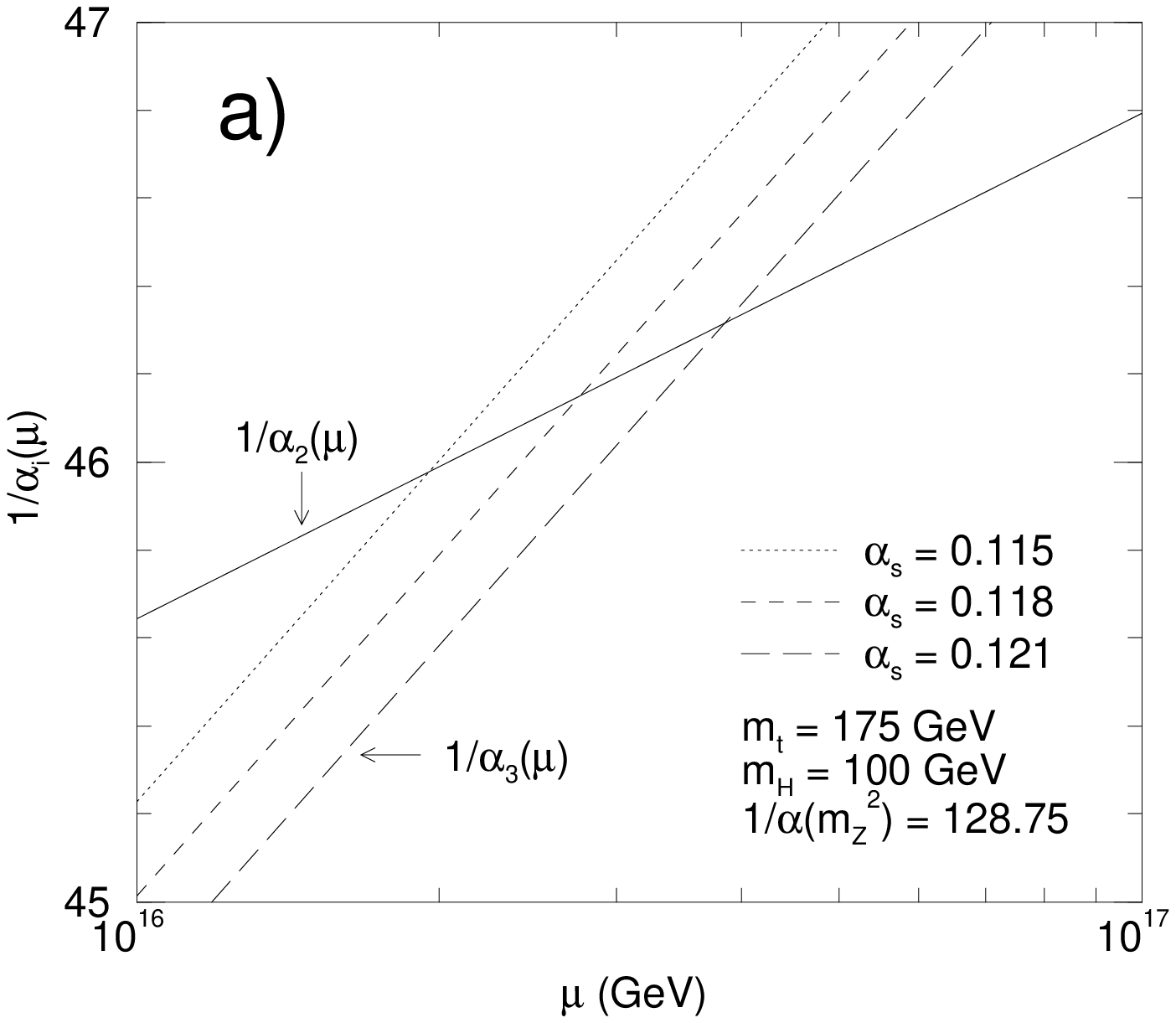,width=6cm}
\leavevmode\psfig{figure=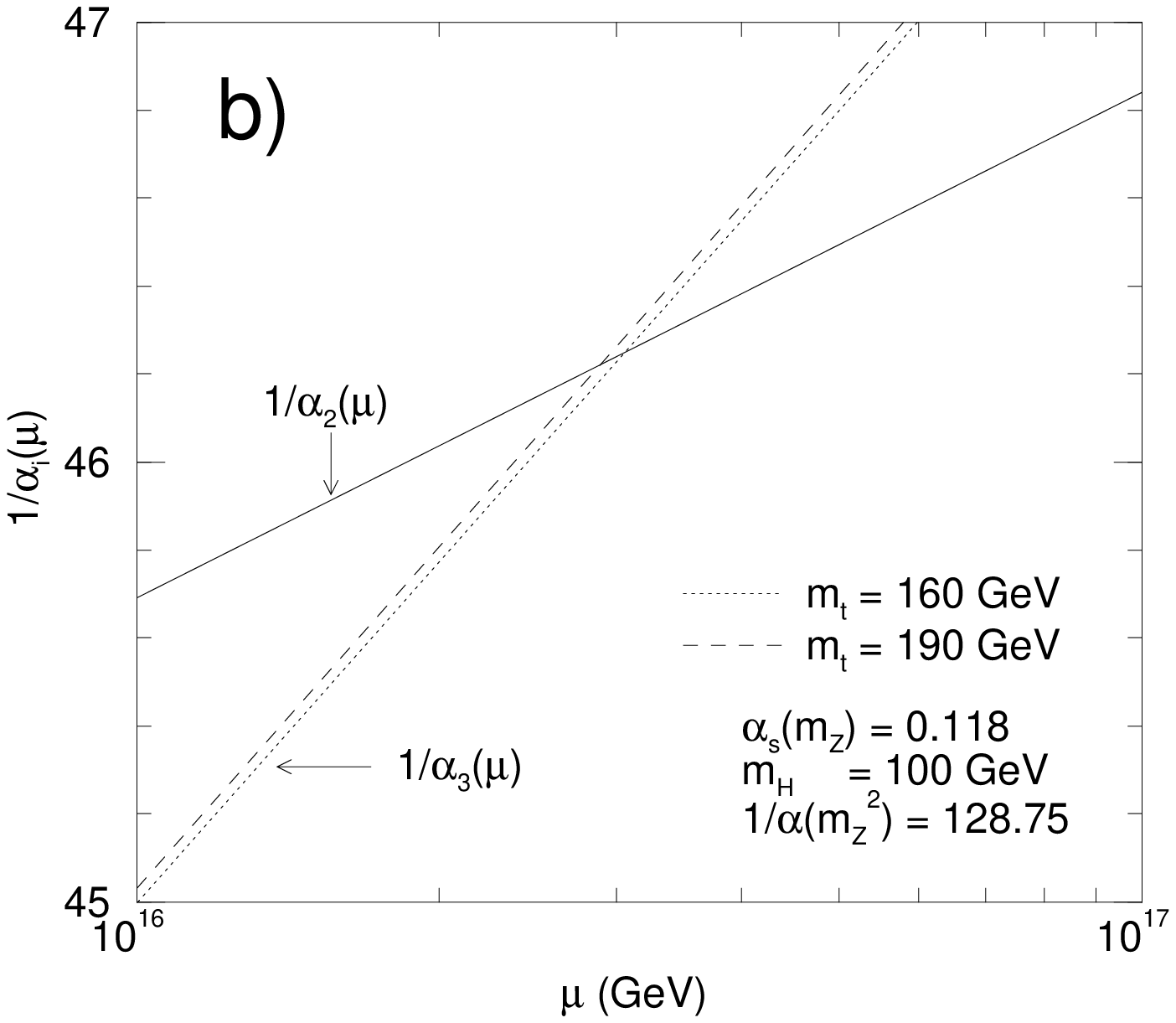,width=6cm}
\leavevmode\psfig{figure=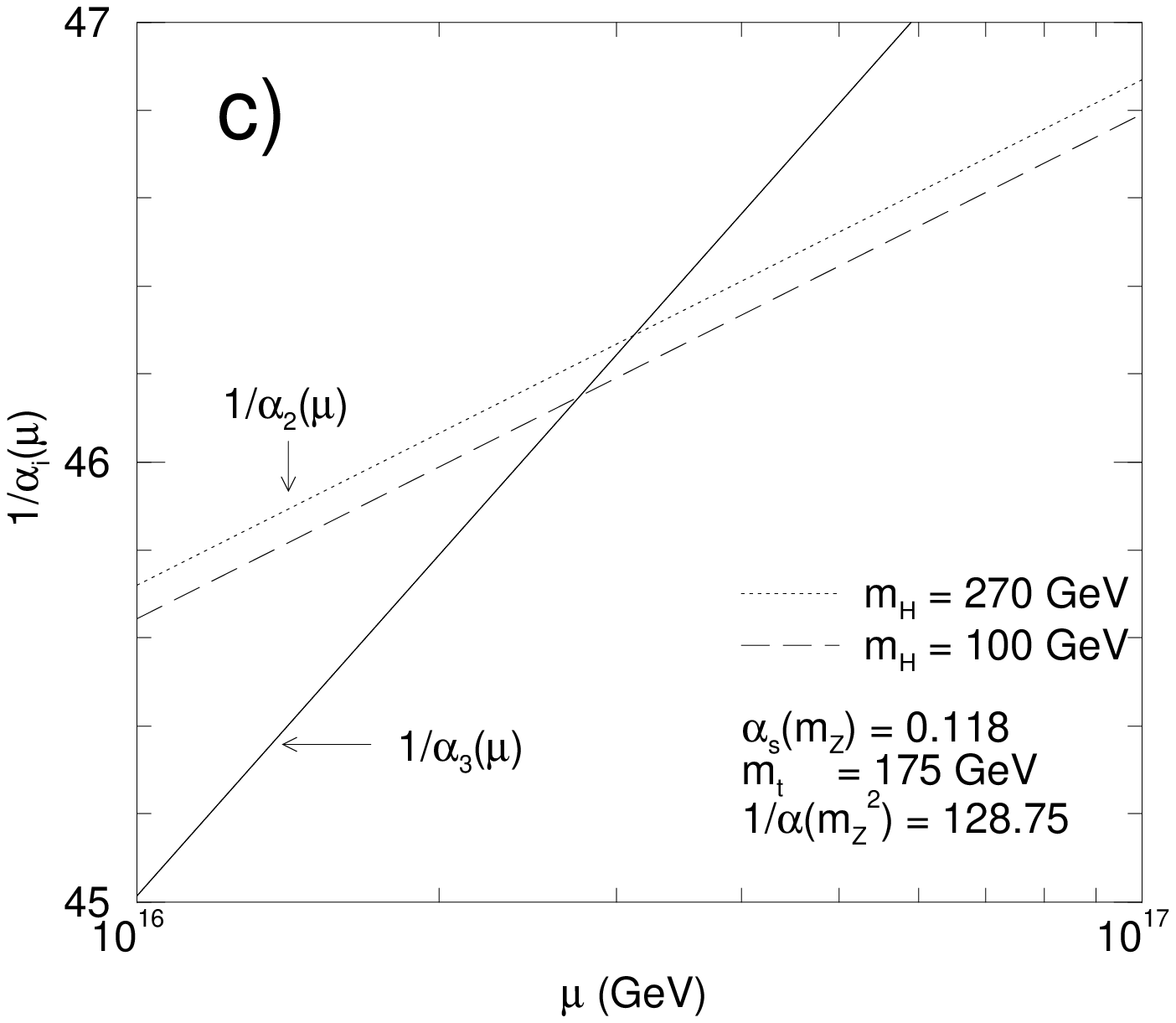,width=6cm}
\leavevmode\psfig{figure=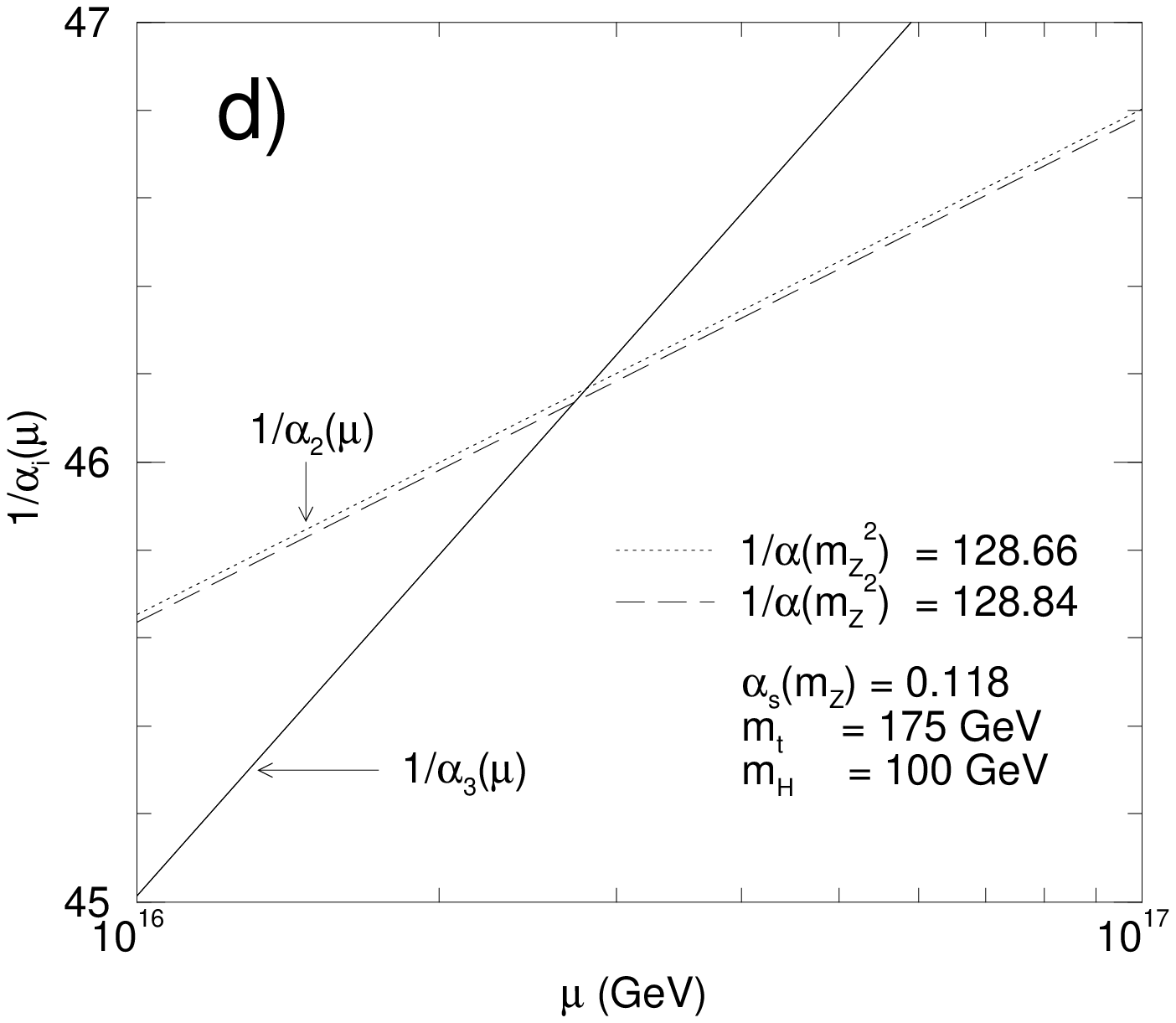,width=6cm}
\end{center}
\caption{
Four parameter dependences of 
the SU(2)$_L$ and SU(3)$_C$ running couplings. 
Each figures correspond to: 
a) $\alpha_s(\mz) = 0.118 \pm 0.003$, 
b) $160~\gev < \mt < 190~\gev$, 
c) $100~\gev < \mh < 270~\gev$, 
d) $1/\abar(m_Z^2) = 128.75 \pm 0.09.$ 
 }
\label{dep_4par}
\vsp{0.5}
\end{figure}

We show the result of our numerical study in Fig.~\ref{dep_4par}. 
In order to show the $\alpha_s(\mz)$-dependence explicitly, 
we choose $\mt = 175~\gev, \mh = 100~\gev$ and 
$1/\abar(m_Z^2) = 128.75$ in Fig.~\ref{dep_4par}a.
In the other figures, we fixed 
$\alpha_s(\mz) = 0.118$ in Figs.~\ref{dep_4par}b, \ref{dep_4par}c 
and \ref{dep_4par}d, 
$\mt = 175~\gev$ in Figs.~\ref{dep_4par}c and \ref{dep_4par}d, 
$\mh = 100~\gev$ in Figs.~\ref{dep_4par}b and \ref{dep_4par}d, 
and $1/\abar(m_Z^2) = 128.75$ in Figs.~\ref{dep_4par}b and 
\ref{dep_4par}c. 
%
%
From Fig.~\ref{dep_4par}, it is clearly seen that 
the 2-loop RGE gives the unification scale $\munif$ which is 
much smaller than the 1-loop RGE 
estimate of eq.~(\ref{mu_sm_1loop}). 
The scale $\munif$ increases for larger $\alpha_s(\mz)$, 
larger $\abar(m_Z^2)$, larger $\mh$, and for smaller $\mt$. 
We find the following parametrization:
\bsub
\bea
\munif
&=& 2.75 + 0.93 \xs + 0.13 \xs^2 -0.20 \xt + 0.30 \xh + 0.03 x_H^2 
\nonumber \\
&& ~~~~~~~- 0.04 \xa~~~~~(\times 10^{16}~{\rm GeV}) ,
\label{eq:munif}\\
\vsk{0.3}
\au^{-1} &=&  46.15 + 0.16 \xs -0.07 \xt + 0.12 \xh + 0.004 x_H^2
	- 0.02 \xa, 
\eea
\label{eq:parametrize}
\esub 
for the unification scale $\munif$ and the unified coupling $\au$.
It is remarkable that the unification scale of the minimal SM as 
determined above is almost the same as that of the MSSM, 
eq.~(\ref{eq:mx_mssm}). 
We can find from eq.~(\ref{eq:munif}) that the largest 
value of the unification scale is $\munif \sim 4.4 \times 10^{16}~\gev$
for $\alpha_s(\mz) = 0.121, \mt = 165~\gev, \mh = 270~\gev$ and 
$1/\abar(m_Z^2) = 128.66$. 
Even with the extreme choice of $\alpha_s(\mz) = 0.124~(\xs = 2)$, 
the scale can reach $\munif \sim 5.7 \times 10^{16}~\gev$. 
It is still smaller than the expected string scale about one order 
of magnitude.
%
%
\begin{figure}[t]
\begin{center}
\leavevmode\psfig{figure=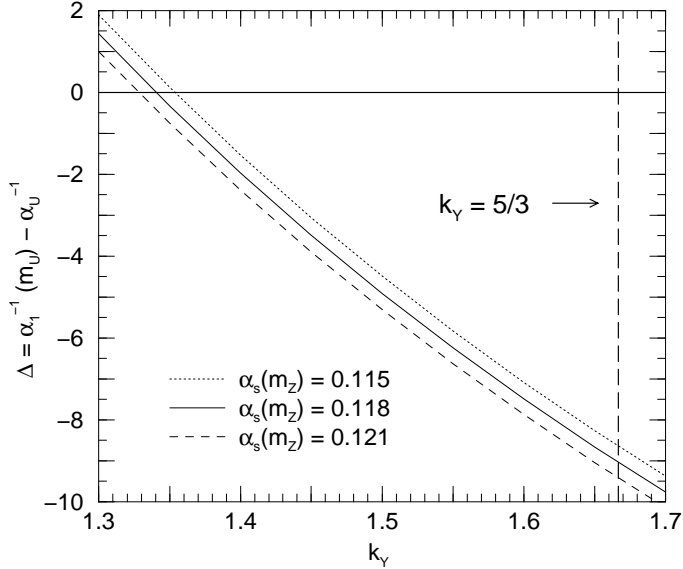,width=9cm}
\end{center}
\caption{
Parameter $\Delta$ as a function of the hyper-charge 
Kac-Moody level $\ky$ for $\mt = 175~\gev, \mh = 100~\gev$, 
and $1/\abar(\mmz) = 128.75$.
The desired $\ky$ is given at $\Delta = 0$ where 
the three gauge couplings are unified. }
\label{ky_delta}
\vsp{0.5}
\end{figure}

The above result tells us that the string unification 
requires either extra matter particles or non-perturbative 
effects, as discussed in ref.~\cite{witten}, even in the 
non-SUSY minimal SM. 
There may also be a possibility that the 2-loop string effects 
can lower the unification scale. 
The desired Kac-Moody level $\ky$ is then found by studying 
the difference 
\beq
\disp{\Delta \equiv 1/\ahat_1(\munif) 
- 1/\alpha_U}, 
\label{Delta}
\eeq
where $\ahat_1(\mu) = \ky \ahat_Y(\mu)$. 
In the absence of the significant string threshold corrections 
among the gauge couplings, the desired range of $\ky$ that gives 
the unification of all three gauge couplings is determined by 
the condition $\Delta = 0$. 
We show $\Delta$ as a function of $\ky$ in Fig.~\ref{ky_delta} 
for $\mt=175~\gev$, $\mh=100$ GeV and $1/\abar(m_Z^2) = 128.75$.
We find that the unification is achieved when 
$1.33~\lsim~\ky~\lsim~1.35$ for $\alpha_s(\mz) = 0.118 \pm 0.003$.
On the other hand, the SU(5) case, $\ky = 5/3$, gives 
$\Delta = -9.02 \pm 0.38$. 

To summarize, 
we have quantitatively studied the possibility of the gauge coupling 
unification of the minimal non-SUSY SM at the string scale with a 
non-standard Kac-Moody level $\ky$. 
Taking into account of the threshold corrections at the boundary of 
the RGE given by $\mt$ and $\mh$, and the uncertainties in 
$\alpha_s(\mz)$ and $\abar(m_Z^2)$, we calculated the unification 
scale $\munif$ in the next-to-leading order. 
Current theoretical and experimental knowledge 
then tells us that the unification scale should satisfy 
$\munif~\lsim~4\times10^{16}~\gev$, which is one order of magnitude 
smaller than the naive string scale of 
$5 \times 10^{17}~\gev$~\cite{string_unif, dienes}. 
If non-perturbative string effects or perturbative higher order effects  
can lower the string scale, then 
the hyper-charge Kac-Moody level should be $1.33~\lsim~\ky~\lsim~1.35$. 
\vsp{1}

We thank H. Aoki and H. Kawai for fruitful discussions. 
The work of G.C.C. is supported in part by Grant-in-Aid for Scientific 
Research from the Ministry of Education, Science and Culture of Japan.

\newpage

\end{document}